\documentclass[prl,reprint]{revtex4-1}
\usepackage{ae}
\usepackage[pdftex]{graphicx}
\usepackage{amsmath}
\usepackage{amssymb}
\usepackage{epsfig,float,color}
\usepackage{hyperref}
\usepackage{color}
\usepackage[dvipsnames]{xcolor}
\usepackage{mathtools}

\begin{document}
\date{\today }

\title{Online Group Dynamics Reveal New Gel Science}
\author{Pedro D. Manrique$^1$, Sara El Oud$^2$, Neil F. Johnson$^{2,*}$}
\affiliation{$^{1}$Theoretical Biology and Biophysics Group, Los Alamos National Laboratory, Los Alamos, NM 87545, U.S.A.}
\affiliation{$^{2}$Physics Department, George Washington University, Washington D.C. 20052, U.S.A.}
\affiliation{$^{*}$neiljohnson@gwu.edu}

\begin{abstract} 
{\bf A better understanding of how support evolves online for undesirable behaviors such as extremism and hate, could help mitigate future harms \cite{GW,3,Gill1,Gill2,shapiro1,NYT,boogs,G1,Gill3,1,2,4}. Here we show how the highly irregular growth curves of groups  supporting two high-profile extremism movements, can be accurately described if we generalize existing gelation models to account for the facts that the number of potential recruits is time-dependent and humans are heterogeneous \cite{TimHH,c1,c2,c3,c4,Palla07,RednerBook,Stockmayer43,Benz01,Flory53,Anita,Ziff82,Lushnikov06,Gillespie77,Newman,Gillespie76,5}. This leads to a novel generalized Burgers equation that describes these groups' temporal evolution, and predicts a critical influx rate for potential recruits beyond which such groups will not form. Our findings offer a new approach to managing undesirable groups online -- and more broadly,  managing the sudden appearance and growth of large macroscopic aggregates in a complex system -- by manipulating their onset and engineering their growth curves.}
\end{abstract}
\maketitle

\noindent {\bf Introduction}

Theories of aggregation have had a successful history in physics, chemistry and beyond \cite{TimHH,c1,c2,c3,c4,Palla07,RednerBook,Stockmayer43,Benz01,Flory53,Anita,Ziff82,Lushnikov06,Gillespie77,Newman,Gillespie76,5}. Understandably, most models and analyses in physical and chemical systems have considered constant size populations of $N$ identical objects in a constant volume space. The resulting aggregation can lead to the sudden appearance of a macroscopically large aggregate, i.e. a gel. Network science has reinforced the importance of this gel formation problem in the context of the growth of a giant connected component (GCC) in a network \cite{RednerBook,Newman}. Indeed, gel and GCC equations can be identical when the system is treated in an averaged way (i.e. mean-field)\cite{RednerBook,Newman}.

Though unrelated in topic, society is currently experiencing another example of something large appearing suddenly `out of nothing', in terms of the rapid recent rise of extremism and hate on social media. The question of {\em why} these undesirable behaviors arise has inspired remarkably in-depth studies across the social sciences \cite{GW,3,Gill1,Gill2,shapiro1,NYT,boogs,G1,Gill3,1,2,4,5} (see Supplementary Information SI for fuller list of references). 
However, social media companies and governments are still struggling with the operational question of {\em how} such undesirable behavior manages to suddenly appear out of nowhere on their platforms and grow so quickly, and what can be done to prevent it.

Here we introduce a generalized aggregation model (Fig. 1) whose solutions can explain how undesirable behavior such as extremism grows online (Fig. 2) and which offers a new approach to mitigation and prevention (Figs. 3 and 4). By incorporating the complexities of the online world that the online user population is {\em time-dependent} ($N(t)$) and individuals are {\em non-identical} and operate in an expansive online space (Fig. 1(a)), the model also happens to represent a novel generalization of existing models in physics and chemistry. Hence our findings also make the broader contribution of advancing understanding of non-equilibrium phenomena in open systems \cite{prl18,Char1,Char2,Char3,arxiv}. We also note that even for the 
\begin{figure}[H]
\centering
\includegraphics[width=0.9\linewidth]{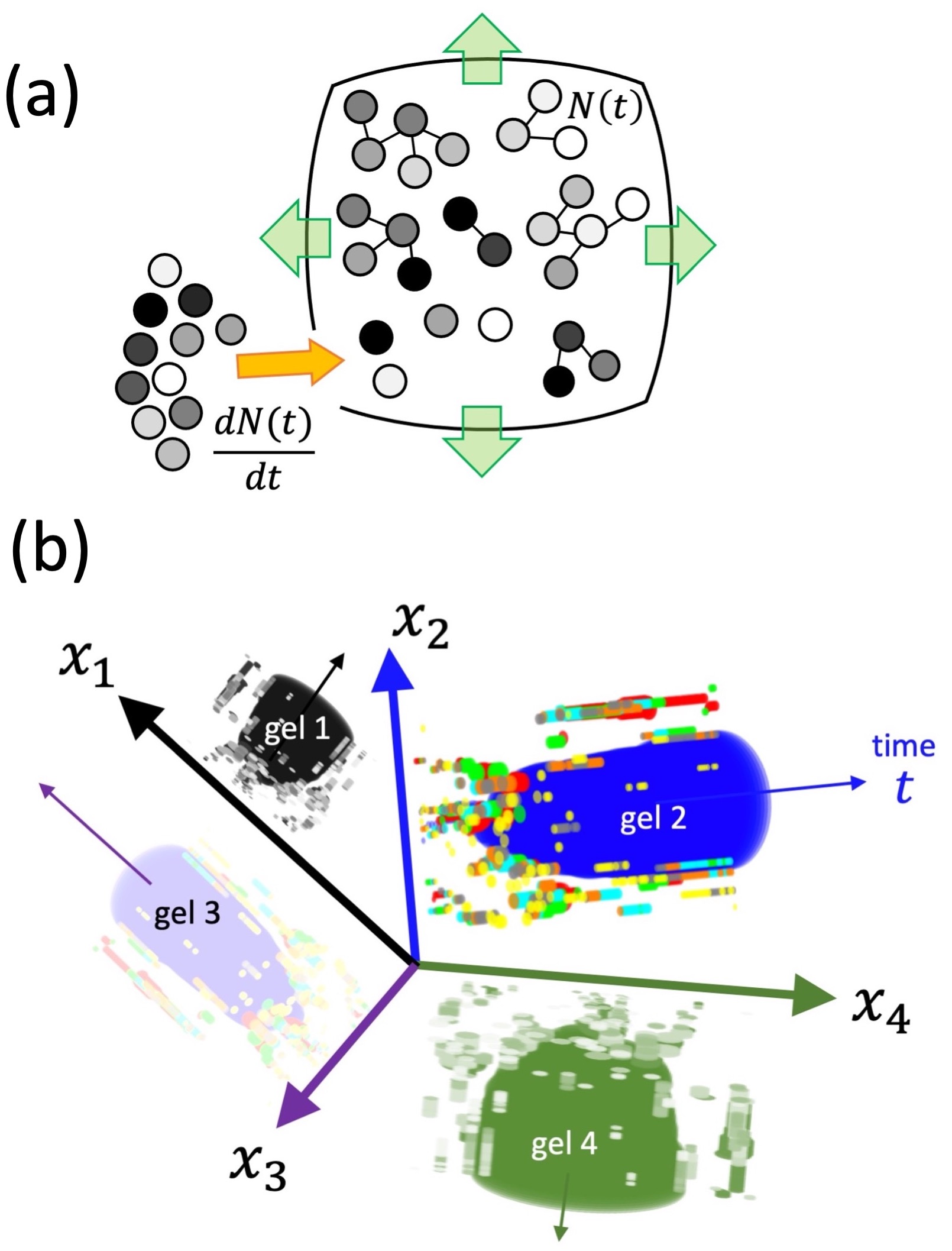}
\caption{\small{(a) $N(t)$ heterogeneous individuals  can aggregate over time. Each individual $i$ has its own internal variable (shade of gray) $\vec{x_i}=(x_{1;i},x_{2;i},\dots x_{T;i})$. (b) Each gel (or equivalently GCC using a network picture) is an observable online group in Fig. 2. Each of the $T$ dimensions mimics a trait or stance on an issue, around which an online group (gel, GCC) can form. Since individuals can simultaneously belong to any number of online groups, each axis can in principle form a group from all $N(t)$ individuals. Groups that combine dimensions could form, but this just shifts the numerical value of the population-averaged aggregation probability $F$.}}
\end{figure}
\noindent online world, the interpretation of the model is not limited exclusively to undesirable behaviors such as extremism, and hence could be used to interpret other irregular growth curves such as those reported recently for online Covid-19 (mis)information \cite{Quatt}.

What makes this otherwise general model (Fig. 1) particularly well-suited to describe online extremism  and hate, is that it captures the empirical finding \cite{groups1,groups2,groups3,VKScience16,SciRep18,Neil19} that individuals interested in such unacceptable or otherwise sensitive topics tend to utilize the inbuilt group-creation features provided by social media platforms such as Facebook and its Europe-based clone 
VKontakte (but not Twitter). In short, it describes  how supporters `gel' together into a group along one of the axes in Fig. 1(b), with the benefits that the group provides a fairly shielded online environment and has a flavor of extremism or hate that appeals to them. Such groups are referred to by different names depending on the platform (e.g. Facebook Page, VKontake Club) and we stress they have nothing to do with communities inferred from a network analysis algorithm. 
Furthermore, the model produces group growth curves (Fig. 2) that are far more irregular than for existing models \cite{prl18,Char1,Char2,Char3,arxiv} but which are close to the empirical curves for  two high profile extremist movements. These movements are the Boogaloos in the U.S., who were reportedly involved in the January 2021 Capitol riot and have members drawn from highly diverse ideologies \cite{GW,boogs}; and ISIS (Islamic State). Our data collection methodology is the same as Refs. \cite{Neil19,VKScience16,SciRep18} and avoids requiring any information about individuals: we first search manually for an initial seed of groups, and then we track which other groups they connect to or mention, in order to iterate toward a closed list (see SI for details). 
  
\begin{figure}
\centering
\includegraphics[width=1.0\linewidth]{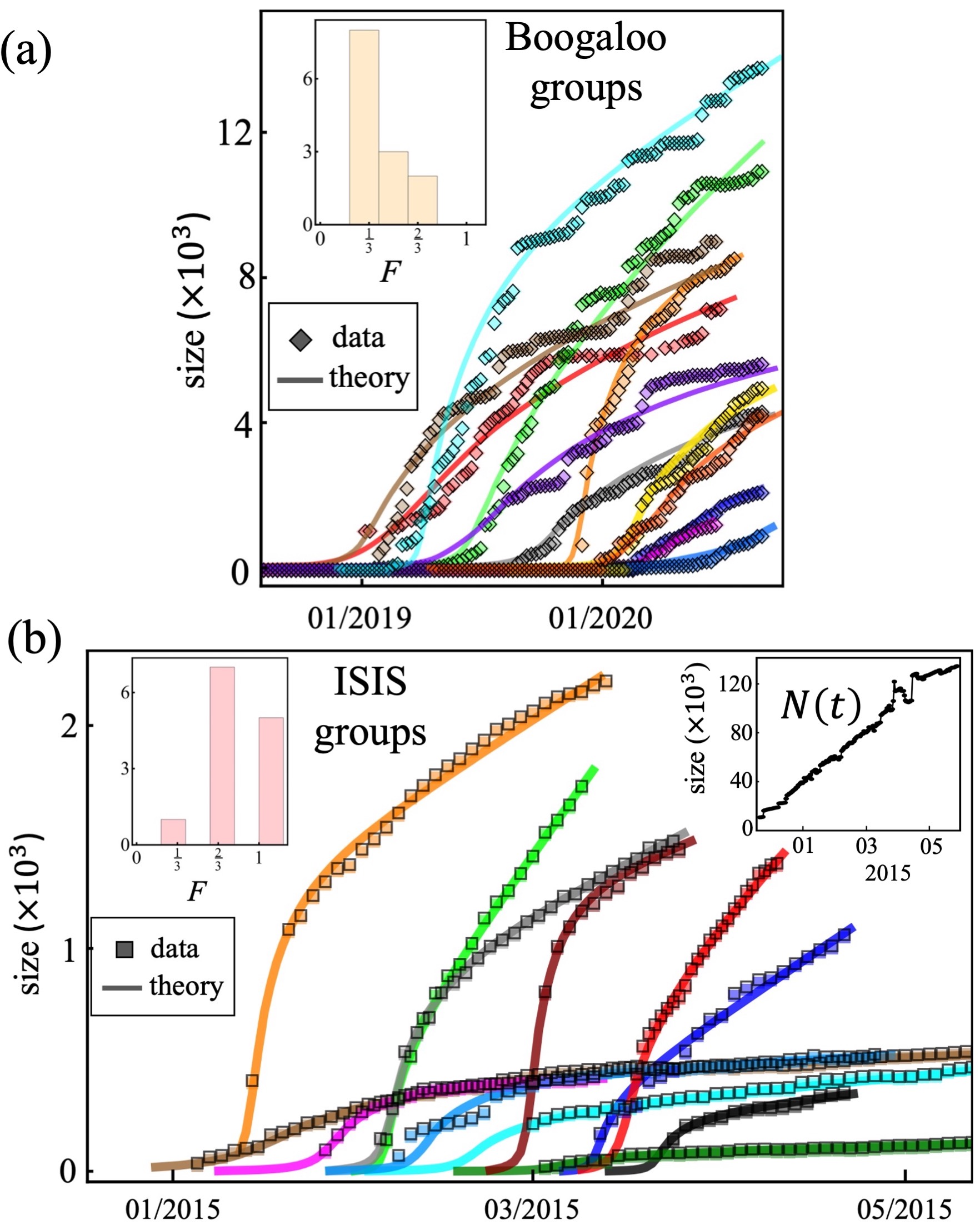}
\caption{\small{Empirical growth of online extremist groups (symbols) compared to numerical solutions of Eq. 2 (curves). For clarity, only a subset of all the empirical groups are shown. See SI for full statistics. Each curve is a separate solution of Eq. 2, and  represents a separate group (gel, or GCC) in Fig. 1(b). The histogram compiling the inferred $F$ value from each of the groups, is shown in the inset. (a) Domestic U.S. extremist groups that identify as Boogaloos, on Facebook. The low $F$ values for the groups ($F\sim 1/3$) are consistent with recent social science case studies \cite{GW} that uncovered high diversity among Boogaloo groups. (b) International U.S. extremist groups that support ISIS, on the less moderated platform VKontakte. The higher $F$ values ($F\sim 2/3$) as compared to (a), are consistent with higher similarity within ISIS supporter groups. Other inset shows $N(t)$ over time. $N(t)$ is not accessible for Facebook in (a). }}
\end{figure}

\noindent {\bf Results}

The model (Fig. 1) starts by assigning each individual $i$ a list of $T$ traits or stances around a given issue,
$\vec{x_i}=(x_{1;i},x_{2;i},\dots,x_{T;i})$ with $i=1,2,\dots N(t)$, drawn from a general heterogeneity distribution $\mathcal{P}\{\vec{x_i}\}$. Centola et al. \cite{Centola} discuss how such a simple approximation is nonetheless consistent with in-depth studies in sociology.  Neither the nature nor number $T$ of these traits or stances affects the form of the equations that we develop -- and they could adapt over time but here are kept constant. For notational simplicity, we consider $T=1$ in what follows. Starting from a population $N(0)$ of isolated individuals, at each timestep we connect two randomly chosen individuals $i$ and $j$ (and hence the aggregates they may belong to in Fig. 1(a)) with a probability per unit time that depends on the pair's similarity $| x_i - x_j |$. For the equations that follow, we do not need to specify a precise function of similarity at the level of individual pairs since this just controls the numerical value of the population-average pairing probability, $F$. As illustration, the SI calculates $F$ values for a uniform distribution $\mathcal{P}\{\vec{x_i}\}$: pairing favoring similarity (homophily) yields $F=2/3$ while dissimilarity (heterophily) yields $F=1/3$. 
This aggregation process can lead to large-scale connectivity transitions over time in any of the $T$ dimensions, producing a group (gel, or GCC) comprising a non-negligible fraction of the population or network. Newcomers are injected as unattached (Fig. 1(a)). 
At mean-field level, the number of aggregates  of size $s$, given by $n_s$, follows these coupled, dynamical nonlinear equations for $s=1$ and $s\ge 2$:
\begin{eqnarray}
\dot{n}_{1}(t)&=&\dot{N}(t)-2F\frac{n_{1}}{N(t)^{2}}\sum_{r=1}^{\infty}rn_{r}\\
\dot{n}_{s}(t)&=&-2F\frac{sn_{s}}{N(t)^2}\sum_{r=1}^{\infty}{rn_{r}}+\frac{F}{N(t)^2}\sum_{r=1}^{s}rn_{r}(s-r)n_{s-r} \ \ .\nonumber 
\end{eqnarray}
Empirical evidence supporting the product kernel form in Eq. (1), comes from studies of human communication and collaboration networks \cite{Palla07}, while its distance independence reflects the global reach of online interactions. As usual, the time $t$ in such averaged equations is not supposed to correspond to a timestep in a full many-particle or network simulation, or real calendar time.

\begin{figure}[H]
\centering
\includegraphics[width=0.75\linewidth]{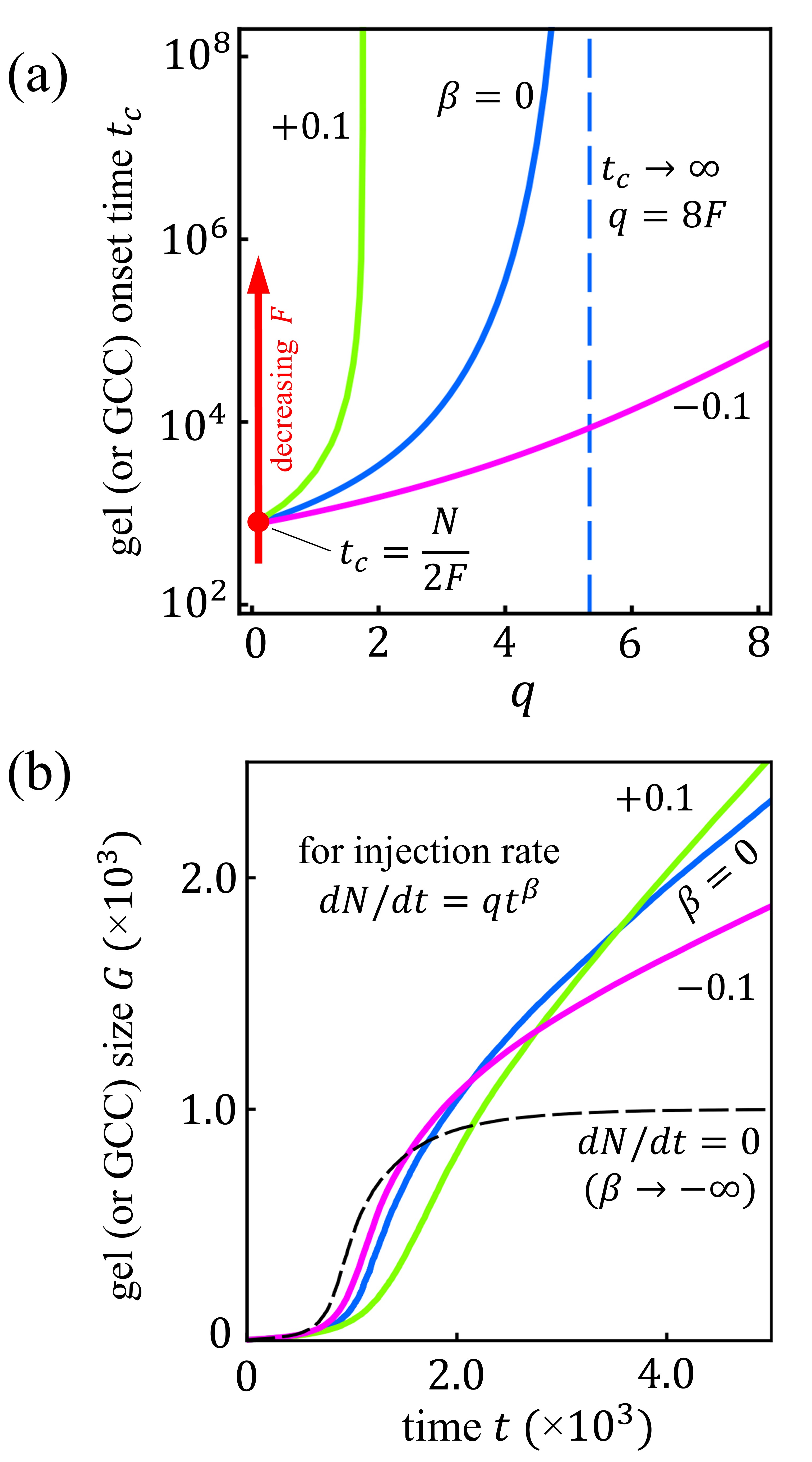}
\caption{\small{Theoretical results with $\dot{N}(t)=qt^\beta$ and different values of $\beta$ for (a) group (gel, GCC) formation time $t_c$ versus $q$. Vertical line for $\beta=0$ shows $q\equiv\dot{N}_{\rm const}|_c=8F$ at which $t_c$ diverges in Eq. 3. Reducing $F$ shifts all the curves upward vertically, and hence makes $t_c$ larger for a given $q$ as shown analytically in Eq. 4 and Fig. 4(a). (b) Group (gel, GCC) time evolution $G(t)$. Calculations are for $N(0)=10^3$, uniform heterogeneity distribution and  $\dot{N(t)}=\dot{N}_{\rm const}=0.5.$
}}
\end{figure}

Using the generating function $\mathcal{E}(y,t)\equiv\sum_{s\geq 1}s n_s e^{ys}$ \cite{prl18,RednerBook} in Eq. 1 (see SI) leads to the following general equation that determines the time-dependent evolution of an online group:
\begin{equation}
\frac{\partial\mathcal{E}(y,t)}{\partial t}+\frac{2F}{N(t)}\left[1-\frac{\mathcal{E}}{N(t)}\right]\frac{\partial\mathcal{E}(y,t)}{\partial y}=e^{y}\frac{dN(t)}{dt} \ .
\label{burgers}
\end{equation}
Setting $\mathcal{E}(0,t)=N(t)-G(t)$ and solving Eq. 2, yields the size of each group over time $G(t)$, as well as its onset time $t_c$ beyond which $G(t)$ becomes non-zero and large. Equation 2 has the form of a generalized Burgers equation with a forcing term. 
We have checked that our numerical, and where possible analytical, solutions of Eqs. 1 and 2 are consistent with full many-particle simulations (i.e. generalized version of the Gillespie algorithm \cite{Gillespie77,Gillespie76}, see SI) and with full network simulations (see SI) that include the full microscopic pairing dynamics and avoid any population-averaging. This helps confirm the suitability of the averaging process for the group (gel) formation and its equivalency to a network picture (GCC).

Equation 2 leads to the conclusion that, in contrast to the usual situation of physical systems with a fixed number $N$ of identical particles in a fixed volume \cite{RednerBook}, the formation of a group (gel, GCC) may be suppressed entirely -- and at the very least can exhibit a wide range of highly irregular growth profiles as shown in Fig. 2.
To demonstrate this explicitly, we adopt a constant  $dN(t)/dt$ (i.e. $\dot N(t)=\dot{N}_{\rm const}$) which is a reasonable approximation for ISIS based on the empirical data in Fig. 2(b) inset, and also for the Boogaloos given the reported steady influx of people during the 2020 pandemic and election buildup. (Individuals could not be counted on Facebook). This leads (see SI) to the exact result for the onset time of a group (gel, GCC):
\begin{equation}
t_{c}=\frac{N(0)}{\dot{N}_{\rm const}}\left[\left(\frac{\alpha-1}{\alpha+1}\right)^{2/\alpha}-1\right]
\label{tc2}
\end{equation}
where $\alpha^2=(1-8F/\dot{N}_{\rm const})$. Equation 3 implies that not only is the group formation time $t_c$ delayed with smaller $F$ and/or larger rate $\dot{N}_{\rm const}$, it can diverge (see Fig. 3(a) for $\beta=0$). As the system gets flooded with more heterogeneous individuals, this slows down the ability for a group (gel, GCC) to emerge -- and can prevent it from ever forming.  This abrupt transition between regimes of eventual group formation and no-group formation, is given by $\dot{N}_{\rm const}|_c=8F$ (vertical line Fig. 3(a)).

This result offers a counterintuitive but novel approach to mitigation: delay the onset of groups (gels, GCCs) by flooding the online space with heterogeneous individuals so that $\dot{N}_{\rm const}>8F$. While such a proposition ultimately requires testing online, it provides a rigorously derived quantitative starting point for policy discussions. For groups that do emerge, their growth can be delayed or flattened by decreasing the $F$ (Figs. 3(a), Fig. 4(a)) which can be achieved by encouraging or engineering more diversity among newcomers. This impact of decreasing $F$ can be seen directly from the following approximate analytic solutions for $G(t)$ shown in Fig. 4(a) and derived in the SI, which are in close quantitative agreement with the simulations:
\begin{eqnarray}
G(t)&=&\frac{N(0)^{\frac{1-\alpha}{2}}N(t)^{\frac{1+\alpha}{2}}}{2\alpha}\left[\alpha-(1+2y_0)\right]+\\
& &\frac{N(0)^{\frac{1+\alpha}{2}}N(t)^{\frac{1-\alpha}{2}}}{2\alpha}\left[\alpha+(1+2y_0)\right]\nonumber\\
&-&\left[\frac{N(t)}{N(0)}\right]^{\frac{1-\alpha}{2}}\frac{z_0}{2\alpha}(\alpha+1)-\left[\frac{N(t)}{N(0)}\right]^{\frac{1+\alpha}{2}}\frac{z_0}{2\alpha}(\alpha-1)\nonumber
\end{eqnarray}
where $y_0=\omega-W(\omega e^{\omega})$, 
$z_0=\frac{N(0)}{\omega}W(\omega e^{\omega})$, $W(.)$ denotes the Lambert function, and
\begin{equation}
\omega=\left[\frac{4F}{\dot{N}_{\rm const}}\right]\left[\frac{N(0)^\alpha-N(t)^\alpha}{N(t)^\alpha(\alpha+1)+N(0)^\alpha(\alpha-1)}\right]\ .
\end{equation}
Independent empirical evidence supporting our finding of a delay in $t_c$ with decreasing $F$, comes from recent laboratory-controlled experiments which find that human groups formed by random aggregation ($F=1$ for uniformly distributed heterogeneity) were quicker to attain a high level of innovation performance than groups formed by like or unlike individuals ($F<1$) \cite{Sayama19}.

\begin{figure}[H]
\centering
\includegraphics[width=1.0\linewidth]{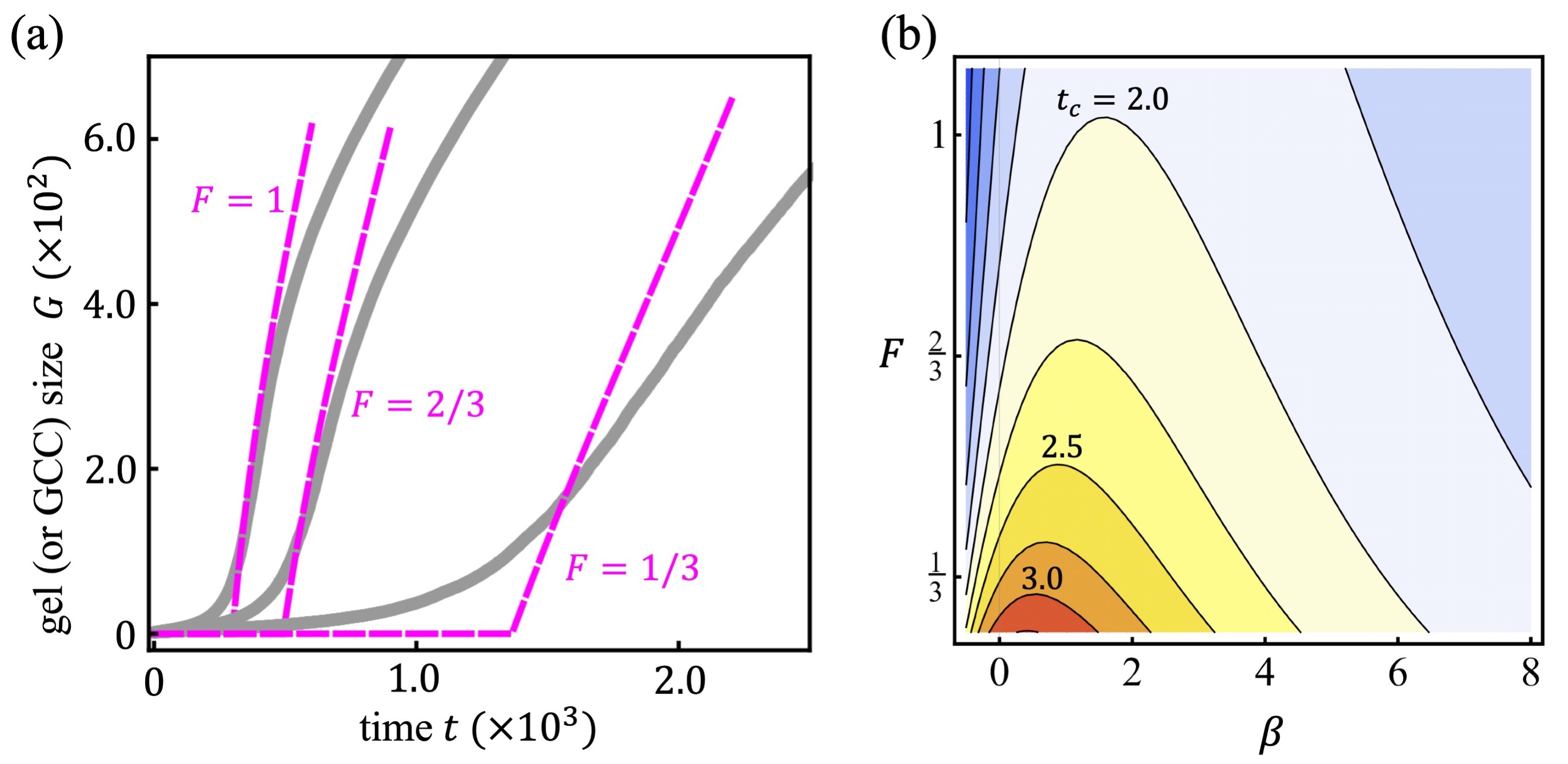}
\caption{\small{(a) Analytic solutions for group (gel, GCC) evolution $G(t)$ (dashed pink curves) from Eq. 4 compared to exact evolution from simulations (solid gray curves averaged over 100 runs). Calculations are for $N(0)=500$, uniform heterogeneity distribution and $\dot{N(t)}=\dot{N}_{\rm const}=0.5$. (b) Example of the rich dependence of $t_c$ on $\beta$ that emerges with  $\dot{N}(t)=qt^\beta$, if the system volume is constrained (see SI) and $N(0)$ is negligibly small.
}}
\end{figure}

Figure 2 compares the numerical solutions of Eq. 2 for $\dot{N}(t)=\dot{N}_{\rm const}$, to the empirical data for the extremist groups. Each group corresponds to a gel (or equivalently GCC) forming along a separate trait (or issue) axis in Fig. 1(b). The rapid growth of these extremism support groups is generally far quicker and irregular than groups focused on benign topics. Hence although our model is not limited to extremism, it again seems particularly appropriate to such undesirable behavior.

Equation 2 can be used as a starting point for quantifying the impact of other policies and interventions, in addition to our suggested one of pushing $\dot{N}(t)$ beyond $8F$ for the constant rate case. For example, suppose the social media platform tries to dynamically control newcomer flow $\dot{N}(t)$. Figure 3 shows the impact for $\dot{N}(t)=qt^\beta$ ($\dot{N}(t)=q\equiv\dot{N}_{\rm const}$ for $\beta=0$). Increasing $\beta$ with $\beta>0$, the onset $t_c$ gets further delayed and the transition to a no-group formation regime occurs at smaller $q$. By contrast, $\beta<0$ appears to remove this transition. Figure 3(b) shows the corresponding group growths $G(t)$. $G(t)$ for $\beta>0$ rises initially more slowly than for $\beta\leq 0$, but eventually overtakes. As $\beta$ becomes increasingly negative, $G(t)$ rises quicker but saturates faster, eventually reaching the constant $N$ (i.e. $\dot{N}(t)=0$) limit that provides a crude initial approximation for some growth curves \cite{prl18,arxiv}. This same $\dot{N}(t)=0$ limit is also reached by setting $q\rightarrow 0$.

Another possible policy would be for platforms to restrict the range of online community spaces that newcomers have access to. We mimic this by fixing the volume in Fig. 1(a) and hence setting the denominators in Eq. 1 to $N(0)$ for all $t$. For $\beta=0$ and hence $\dot{N}(t)=\dot{N}_{\rm const}$, the onset time is:
\begin{equation}
t_c=\left(\frac{\pi}{2}-\arctan{\sqrt{\frac{2F}{\dot{N}_{\rm const}}}}\right)\frac{N(0)}{\sqrt{2F\dot{N}_{\rm const}}}
\end{equation}
which has no transition to a no-group formation regime, i.e. a group always emerges eventually. This onset time is quicker than for the prior discussed cases, including constant $N(t)$.  More generally, for $\beta\neq 0$, $t_c$ satisfies the following transcendental equation:
\begin{equation}
AJ_{\frac{1}{\beta+2}}\left(\frac{2\gamma}{\beta+2}t_c^\frac{\beta+2}{2}\right)=J_{-\frac{1}{\beta+2}}\left(\frac{2\gamma}{\beta+2}t_c^\frac{\beta+2}{2}\right),
\end{equation}
where $J_n$ is the Bessel function of the first kind, and
\begin{eqnarray}
A&=&\frac{2F}{N(0)}\frac{\Gamma(\frac{1}{\beta+2})}{\Gamma(\frac{\beta+1}{\beta+2})}(\beta+2)^{-\frac{\beta}{\beta+2}}\gamma^{-\frac{2}{\beta+2}},\\
\gamma&=&\sqrt{\frac{2F\dot{N}_{\rm const}}{N(0)^{2}}}\ .
\end{eqnarray}
Figure 4(b) illustrates the rich nonlinear dependence that can then arise for $t_c$ as a function of $\beta$ for the limit of $N(0)\rightarrow0$. This and other results for this `constant volume' limit are given in the SI.
 \newpage
\noindent {\bf Discussion}

Our model and analyses are not a priori limited to extremism or human behavior. Hence our findings serve more broadly to indicate  novel science for aggregation in heterogeneous, time-varying populations or networks across the sciences. In particular, our analysis has led to a general equation that describes the onset and time-evolution of macrosopically large groupings (i.e. a macroscopically large correlation or coherence) in a general population which has a time-dependent size and where its component objects are non-identical and operate in an expansive space. This is important since the sudden appearance of such a macrosopically large grouping may be beneficial or harmful to a system (e.g. it may generate extreme events). Our analysis then shows how to manipulate its onset, engineer its growth curve, and even prevent it from forming.

PDM acknowledges support from Los Alamos National Laboratory Director's Fellowship. NFJ acknowledges support of US Air Force Office for Scientific Research through grants FA9550-20-1-0382 and FA9550-20-1-0383. We are grateful to Nicholas Johnson Restrepo, Rhys Leahy, Yonatan Lupu and Nicolas Velasquez for help with the empirical data.

\end{document}